\documentclass{article}
\usepackage{a4wide}

\begin{document}

\newcommand{\ket}[1]{|#1\rangle}
\newcommand{\bra}[1]{\langle#1|}
\newcommand{\scpr}[2]{\langle#1|#2\rangle}

\title{Two-dimensional electron gas in a periodic potential and external 
magnetic field: states of pairs and three-particle systems}

\author{Wojciech Florek\\
{\it A. Mickiewicz University, Institute of Physics}\\
{\it ul. Umultowska 85, 61--614 Pozna\'n, Poland}}
\date{7 Nov, 2000}
\maketitle

\begin{abstract}
 The group-theoretical classification of states 
of identical particle pairs is presented. Then obtained states are coupled 
with those of an antiparticle to construct states of a three-particle 
system. 
Investigations are performed using 
 products of irreducible projective representations of the 2D translation
 group. For a given BvK period $N$ degeneracy of pair states is $N$, whereas 
three-particle states are $N^2$-fold degenerated. It has to be underlined 
that the case of even $N$ is more complicated since pair states are
labelled by four inequivalent irreducible projective representations.
The problem of symmetry properties with respect to particles transposition
is briefly discussed 
\end{abstract}
\noindent{PACS numbers: 73.20.Dx, 71.70.Di, 02.20.-a}

\section{Introduction}

The quantum Hall effect and high temperature superconductivity have 
raised interest in properties of the two-dimensional electron gas subjected to 
electric and
magnetic fields. The observation of (negatively) charged excitons 
\cite{shields} has recalled a forty-year old concept of excitons, `trions'
or `charged excitons', introduced by Lampert in 1958 \cite{lam}.
Recently, such excitons, consisting of two holes and an electron or two 
electrons and hole (denoted $X^\pm$, respectively), have been investigated 
both experimentally and theoretically \cite{exper,theor,Dzy}. In the ealier
paper \cite{trio} the trion states have been discussed; they have been
classified using a direct coupling of three one-particle states and
in a two-step procedure: (i) coupling of particle and antiparticle states
to obtain states of an electrically neutral system and then (ii) coupling
with hole (electron) states to determine states of a trion $X^{\pm}$. 
The aim of this work is to present classification of three-particle
states (strictly speaking states of particle-particle-antiparticle systems)
when at first step states of a pair of indentical (charged) particles
are constructed. The antiparticle states are taken into account in the
second step. Such an approach allow to discuss pair states, what may 
be important in considerations of high-$T_{\mathrm{c}}$ superconductors,
where Cooper pairs are confined to Cu-O planes.

The classification presented in this paper, like in the previous 
one \cite{trio}, is based on translational symmetry in the presence 
of a periodic potential and an external, constant and homogeneous, 
magnetic field. To 
perform this task the so-called magnetic translation operators, introduced by 
Brown \cite{brow} and Zak \cite{zak}, are used. These operators commute with 
the standard Hamiltonian of an electron in the magnetic field 
${\mathbf H}=\nabla\times{\mathbf A}$ and a periodic potential 
$V({\mathbf r})$ 
\begin{equation}\label{ham}
  {\cal H} =
  {1\over{2m}}\left({\mathbf p}+{e\over c}{\mathbf A}\right)^2
     + V({\mathbf r})\,.
\end{equation}
The Born--von K\'arm\'an (BvK) periodic conditions imposed on a system lead
to a finite-dimensional projective representation of the 2D translation group
formed by the magnetic translation operators. The Kronecker products of 
irreducible 
projective representations can be applied to description of multi-particle 
states \cite{flo99,flo99a}, what has been done lately for trions \cite{trio}. 
It seems that investiagtions of pairs should be much easier. 
However, trions, consisting of two electrons
and a hole or two holes and an electron, have the total charge $\pm e$ and,
therefore, behave in a similar way as a single electron and a hole, 
respectively. On the other hand, a pair of electrons (holes) has the 
charge $\mp 2e$ and one has to take into account parity of the 
Born--von K\'arm\'an period $N$. It means that cases of odd and even 
period should be considered separtely. 

Investigating problems, which involve the magnetic field ${\mathbf H}$ 
determined
by the vector potential ${\mathbf A}$, one has to keep in mind that some
results may depend on a chosen gauge, though physical properties should
be gauge-independent. Two gauges are most frequently used in description of 
the 2D electron systems: the Landau gauge with ${\mathbf A}=[0,xH,0]$ and the
antisymmetric one with ${\mathbf A}=({\mathbf H}\times{\mathbf r})/2$. The 
relations between these gauges was discussed in the earlier article 
\cite{jmp}, so this problem is left out in the present considerations. 
However, 
it should be underlined that a form representation matrices depend on chosen 
gauge and, moreover, obtained representations are inequivalent, what means,
among others, that their bases are not related by a unitary transformation.
Since it is a symmetry adapted basis what results from presented material, 
then it is important to stress that the Landau gauge is assumed and obtained
results can not be immediately applied to other gauges. 
For the sake of simplicity the 
considerations are limited to this case and the presented results correspond
to the limit of high magnetic fields, {\it i.e.} there is no Landau level
mixing. 

\section{Pairs of identical particles}

The movement of two-dimensional charged particles in a periodic potential
and, like in the Hall 
effect, subjected to a magnetic field is described by the magnetic 
translation operators introduced by Brown \cite{brow} and Zak \cite{zak}.
These operators form a projective representation of the two-dimensional
translation group $T=\{\mathbf{R}=[n_1,n_2]\mid n_1,n_2 \in {\mathbf Z}\}$ 
\cite{zak};
a series of articles on such representations was 
presented by Backhouse and Bradley \cite{babr}. Possible applications
to multiparticle states were presented by the author in some earlier
works \cite{flo99,flomp1}, where products of projective representations 
has been considered, but the irreducible basis has not been determined.
The first attempt to determine trion states was based on two approaches:
(i) a direct coupling of three one-particle states and (ii) construction 
of particle-antiparticle states and then coupling with one-particle state
\cite{trio}.

Investigating movement of charged particles in a magnetic field one
has to distinguish the periodic (Born--von K\'arm\'an) conditions related
to the periodicty of a crystal lattice and the magnetically periodic
conditions introduced by Brown \cite{brow,jmp}. Let $D$ be a projective
representattion of $T$ with basis vectors labelled by $\ket{w}$. Then, 
in short, the magnetic period $N_{\mathrm{mag}}$ is determined by the 
equality $D(N_{\mathrm{mag}}\mathbf{R})\ket{w}=
 \mathrm{e}^{\mathrm{i}\varphi}\ket{w}$, whereas the Born--von K\'arm\'an
period $N$ requires $\mathrm{e}^{\mathrm{i}\varphi}=1$. It means, 
for example, that standard (one-dimensional) irreducible vector 
representations $\Gamma^{k_1,k_2}[n_1,n_2]\ket{w}=\exp[2\pi\mathrm{i}
(k_1n_1+k_2n_2)/N]\ket{w}$ corresponds to $N_{\mathrm{mag}}=1$.
To simplify the present considerations it is assumed that the magnetic field 
and the Born--von 
K\'arm\'an period $N$ are such that a single particle states are related
to an $N$-dimensional projective representation ({\it i.e.}\
$N_{\mathrm{mag}}=N$)
\begin{equation}\label{djk}
  D^l_{jk}[n_1,n_2]=\delta_{j,k-n_2}\omega_N^{ln_1j}\,,
\end{equation}
 where $l$ is mutually prime with $N$, $\omega_N=\exp(2\pi{\mathrm i}/N)$, 
$j,k,n_1,n_2=0,1,\dots,N-1$ and all expressions are calculated modulo $N$
\cite{jmp}.
This formula can be also expressed by the action of $D^l$ on basis vectors 
$\ket{s}$, $s=0,1,2,\dots,N-1$
  \begin{equation}\label{bvec1}
    D^l[n_1,n_2]\ket{s}
   = \omega_N^{ln_1(s-n_2)} \ket{s-n_2}\,.
  \end{equation}
 It is assumed that a single particle under question is a hole (or any
particle with the charge $q>0$) but this assumption does not lead to
any lost of generality.

\subsection{Representations with odd dimension}
 The case of $N$ odd is very simple. Since $\gcd(l,N)=1$ and $N$ is odd then
$\gcd(2l,N)=1$, too. Therefore ({\it cf.} \cite{flo99,flo99a})
 \begin{equation}\label{oddprod}
    D^l\otimes D^l = N D^{2l}\,,
 \end{equation}
 {\it i.e.} two-particle states are $N$-fold degenearted and related to
the projective representation $D^{2l}$. One of possible forms of the 
irreducible basis is
 \begin{equation}\label{oddbas}
  \ket{s}_2^r = \ket{s+r} \ket{s-r}\,,\qquad 
  s,r=0,1,2,\dots,N-1\,,
 \end{equation}
 where the subscript `2' indicates states of a pair and $r$ is the repetition 
index. For the sake of simplicity, here and thereafter the symbol `$\otimes$' 
is omitted in the tensor product of vectors. This result can be
easily verified using Eq.~(\ref{bvec1}):
 \begin{eqnarray*}
  (D^l\otimes D^l)[n_1,n_2]\ket{s}_2^r
 &=&D^l[n_1,n_2]\ket{s+r}\; D^l[n_1,n_2]\ket{s-r} \\
 &=&\omega_N^{2ln_1(s-n_2)} \ket{s+r-n_2} \ket{s-r-n_2}
 =D^{2l}[n_1,n_2]\ket{s}_2^r\,.
 \end{eqnarray*}

\subsection{Representations with even dimension}
 The case $N=2M$ leads to the representations product in the
following form \cite{flo99}
 \begin{equation}\label{evenprod}
    D^l\otimes D^l 
 = M (D^{l;00}\oplus D^{l;10}\oplus D^{l;01}\oplus D^{l;11})\,,
 \end{equation}
 where each of four representations on the RHS is $M$-dimensional and
they form a complete set of  
inequivalent irreducible projective representations of
${\mathbf Z}_N\otimes {\mathbf Z}_N$ (or, in the other words, 
the magnetic translation operators with $N_{\mathrm{mag}}=M$ 
and the Born--von K\'arman period $N$). The first 
representation, labelled by `$00$', is analogous to this given by 
Eqs.~(\ref{djk}) and (\ref{bvec1})
 \begin{eqnarray}
  D^l_{jk}[n_1,n_2]&=&\delta_{j,k-\eta_2}\omega_M^{l\eta_1j}\,;\label{djkM}\\
  D^l[n_1,n_2]\ket{s}&=&\omega_M^{l\eta_1(s-\eta_2)} 
  \ket{s-\eta_2}\,,\label{bvecM}
 \end{eqnarray}
 where $s=0,1,2,\dots,M-1$ now and $\eta_i=n_i\bmod M$, $i=1,2$. All 
representations can be expressed by a general formula \cite{babr}
 \begin{equation}\label{Mdim}
  D^{l;k_1k_2}[n_1,n_2]\ket{s}
  =\omega_N^{ln_1(2s-2\eta_2+k_1)}(-1)^\alpha\ket{s-\eta_2}\,,
 \end{equation} 
 where $k_1,k_2=0,1$ and $\alpha=1$ for $1\leq n_2-s\leq M$ and $\alpha=0$
otherwise. The irreducible bases can be chosen in the following way 
(vectors are not normalized throughout this paper)
 \begin{eqnarray} 
  \ket{s}_2^{r;00} &=& \ket{s+r}\ket{s-r}
      +\ket{M+s+r}\ket{M+s-r}\,,\label{ba00}\\
  \ket{s}_2^{r;10} &=& \ket{s+r}\ket{s-r+1}
       +\ket{M+s+r}\ket{M+s-r+1}\,,\label{ba10}\\
  \ket{s}_2^{r;01} &=& \ket{s+r}\ket{s-r}
       -\ket{M+s+r}\ket{M+s-r}\,,\label{ba01}\\
  \ket{s}_2^{r;11} &=& \ket{s+r}\ket{s-r+1}
       -\ket{M+s+r}\ket{M+s-r+1}\,,\label{ba11}
 \end{eqnarray} 
   where $s,r = 0,1,2,\dots,M-1$. These formulae can be verified by 
simple calculations.

\section{Trion states}\label{direct}

In the presented approach trion states are constructed from states
of a pair, discused in the previous section, and antiparticle states
related to the $N$-dimensional representation $D^{-l}$. Therefore,
we start from investigation of products $D^{2l}\otimes D^{-l}$ and
$D^{l;k_1k_2}\otimes D^{-l}$, for odd and even $N$, respectively.
In both cases, the resultant representations is a multiplicity of
$D^l$ \cite{trio}, namely
  \begin{eqnarray}
   D^{2l}\otimes D^{-l}&=& N D^l\,, \label{oddtrion}\\
   D^{l;k_1k_2}\otimes D^{-l} &=& M D^l\,. \label{eventrion}
  \end{eqnarray}
   
 The first case ($N$ odd) is again very simple. Let $\ket{w}_t$ 
denote a trion state and $\ket{u}_-$ --- a state of an antiparticle, then 
one of possible choice of basis vectors is as follows
  $$
   \ket{w}_t^v=\ket{w+v}_2 \ket{w+2v}_-\,,\qquad
   w=0,1,2,\dots,N-1,
  $$
  where $v=0,1,2,\dots,N-1$ is the repetition index. Since pair states
have additional repetition index $r$, then the trion states in this case
are labelled by $w,r,v=0,1,2,\dots,N-1$, where $w$ is a vector index and
$r,v$ are repetition indices. Taking into account Eq.~(\ref{oddbas})
one obtains
 \begin{equation}\label{trioddbas}
  \ket{w}_t^{r,v}=\ket{w+v}_2^r\ket{w+2v}_-
  = \ket{w+v+r} \ket{w+v-r} \ket{w+2v}_-\,.
 \end{equation} 
  The states
 \begin{equation}\label{wpq}
    \ket{w}^{pq}_t =
  \ket{w+p}\, \ket{w+q}\, \ket{w+p+q}_-
  \end{equation}
 obtained previously \cite{trio} correspond
to $p=v+r$ and $q=v-r$ (calculated mod~$N$, of course). Note that for $N=2M$ 
this relations can not be inverted since $v=(p+q)/2$ and $r=(p-q)/2$ have
no solutions mod~$N$ for odd $p+q$ and $p-q$, respectively.

In the case $N=2M$ considerations are a bit more difficult but 
since one knows the final results, states $\ket{w}^{pq}$, then they
may serve as a usefull hint. There are $N^2=4M^2$ different bases labelled by
$p,q=0,1,2,\dots,N-1$, however if $p'=p+M$ and $q'=q+M$ then states
$\ket{w}^{pq}$ and $\ket{w}^{p'q'}$ have the same third element in the
tensor products because (see Eq.~(\ref{wpq}))
 \begin{equation}\label{wpqM}
    \ket{w}^{p'q'} =  
  \ket{w+p+M} \ket{w+q+M} \ket{w+p+q}_-\,.
 \end{equation}
  Therefore, they can be gathered into $2M^2=NM$ pairs
  $$
   \ket{w}^{pq}\qquad \mbox{and}\qquad \ket{w}^{(p+M)(q+M)}\,. 
  $$
  The ranges of indices have to be chosen in such a way that pairs $pq$
and $(p+M)(q+M)$ run over two separte sets. This problem will be discussed
below in more details. For each pair $p,q$ one 
can form two new bases, labelled by `+' and `--', respectively,
  \begin{equation}
    \ket{w}^{pq\pm}=
   \ket{w}^{pq}\pm \ket{w}^{(p+M)(q+M)}\,, 
  \end{equation}
  so
  \begin{equation}\label{newbas}
    \ket{w}^{pq\pm}=
  \left(\ket{w+p} \ket{w+q}\pm
  \ket{w+p+M} \ket{w+q+M}\right) \ket{w+p+q}_-\,.
  \end{equation}
 These vectors have the first part (in parentheses) in the form resembling
equations (\ref{ba00})--(\ref{ba11}). Therefore, one has to relate 
the repetition index $r$ and a label $u$ of a vector $\ket{u}_-$ of the 
representation $D^{-l}$ with the repetition indices $p,q$ and $\pm$ in
Eq.~(\ref{newbas}). Moreover, due to the relation (\ref{eventrion}) the 
repetion index $v=0,1,\dots,M-1$ has to be taken into account. The similarity
of Eq.~(\ref{newbas}) and Eqs.~(\ref{ba00})--(\ref{ba11}) suggests that the 
irreducible basis of the product $D^{l;k_1k_2}\otimes D^{-l}$ can be chosen
as tensor products $\ket{s}_2\ket{u}_-$. The simplest to
solve is the case of representations with $k_2=0$, {\it i.e.}\ 
$D^{l;k_10}$:
 \begin{eqnarray}
  \ket{w}_t^v&=&\ket{s+v}_2^{00}\;\ket{w+2v}_-\,,  \label{rep1}\\
  \ket{w}_t^v&=&\ket{s+v}_2^{10}\;\ket{w+2v+1}_-\,, \label{rep2}
 \end{eqnarray}
 with $s=w\bmod M$; $s+v$ is calculated mod~$M$, whereas $w+2v$ and 
$w+2v+1$ are calculated
 mod~$N$. For example, when $N=6$ and $l=1$ one obtain bases in the 
following form
  \begin{eqnarray*}
  \left\{\ket{0}_2^{00}\ket{0}_-\;, \ket{1}_2^{00}\ket{1}\;,
    \ket{2}_2^{00}\ket{2}_-\;, \ket{0}_2^{00}\ket{3}\;,
      \ket{1}_2^{00}\ket{4}_-\;, \ket{2}_2^{00}\ket{5}\right\}\;;\\
  \left\{\ket{1}_2^{00}\ket{2}_-\;, \ket{2}_2^{00}\ket{3}\;,
    \ket{0}_2^{00}\ket{4}_-\;, \ket{1}_2^{00}\ket{5}\;,
      \ket{2}_2^{00}\ket{0}_-\;, \ket{0}_2^{00}\ket{1}\right\}\;;\\
  \left\{\ket{2}_2^{00}\ket{4}_-\;, \ket{0}_2^{00}\ket{5}\;,
    \ket{1}_2^{00}\ket{0}_-\;, \ket{2}_2^{00}\ket{1}\;,
      \ket{0}_2^{00}\ket{2}_-\;, \ket{1}_2^{00}\ket{3}\right\}\;;\\
  \left\{\ket{0}_2^{10}\ket{1}_-\;, \ket{1}_2^{10}\ket{2}\;,
    \ket{2}_2^{10}\ket{3}_-\;, \ket{0}_2^{10}\ket{4}\;,
      \ket{1}_2^{10}\ket{5}_-\;, \ket{2}_2^{10}\ket{0}\right\}\;;\\
  \left\{\ket{1}_2^{10}\ket{3}_-\;, \ket{2}_2^{10}\ket{4}\;,
    \ket{0}_2^{10}\ket{5}_-\;, \ket{1}_2^{10}\ket{0}\;,
      \ket{2}_2^{10}\ket{1}_-\;, \ket{0}_2^{10}\ket{2}\right\}\;;\\
  \left\{\ket{2}_2^{10}\ket{5}_-\;, \ket{0}_2^{10}\ket{0}\;,
    \ket{1}_2^{10}\ket{1}_-\;, \ket{2}_2^{10}\ket{2}\;,
      \ket{0}_2^{10}\ket{3}_-\;, \ket{1}_2^{10}\ket{4}\right\}\;.
  \end{eqnarray*}
  These results can be used to determine the set a pair of indices $pq$ 
run over.
However, in each case one has two possibilities, $pq$ and $(p+M)(q+M)$, 
and the appropriate choice can be done after considering representations
$D^{l;k_11}$. The Eqs.~(\ref{ba00}) and (\ref{ba01}) with formulae
(\ref{rep1}) and (\ref{rep2}) give us bases 
  $\ket{w}_t^{r,v;00}$ and $\ket{w}_t^{r,v;10}$, respectively
(as above $s=w\bmod M$) 
 \begin{eqnarray}
   \left(\ket{s+r+v}\ket{s-r+v}
      +\ket{M+s+r+v}\ket{M+s-r+v}\right)\;
   \ket{w+2v}_-\,,\label{ba00r1}\\
   \left(\ket{s+r+v}\ket{s-r+v+1}
      +\ket{M+s+r+v}\ket{M+s-r+v+1}\right)\;
   \ket{w+2v+1}_-\,.\label{ba10r2}
 \end{eqnarray}
   Therefore, the admissible values of indices $(pq)$ are
$(v+r)(v-r)$ and $(v+r)(v+1-r)$ for $v,r=0,1,\dots,M-1$ and all values
are calculated mod~$N$. In the considered example $N=6$ one obtaines
the following 18 pairs: 
`$00$, `$11$', `$22$', `$15$', `$20$', `$31$', `$24$', `$35$', `$40$',  
`$01$, `$12$', `$23$', `$10$', `$21$', `$32$', `$25$', `$30$', `$41$'.  
  In a general case the two following arrays can be constructed:
$$
\begin{array}{c|ccccc}
\multicolumn{1}{c|}{\raisebox{-6pt}{$v$}\quad\raisebox{1pt}{$r$}} & 
     0 & 1 & 2 & \dots &  M-1 \\ \hline
 0 & 00 & 1(N-1) & 2(N-2) & \dots & (M-1)(M+1)  \\
 1 & 11 & 20  & 3(N-1) & \dots & M(M+2)  \\
 2 & 22 & 31 & 40 & \dots & (M+1)(M+3) \\
 \dots & \dots & \dots & \dots  & \dots \\
 M-1 & (M-1)(M-1) & M(M-2) & (M+1)(M-3) & \dots & (N-2)0 
\end{array}
$$
$$
\begin{array}{c|ccccc}
\multicolumn{1}{c|}{\raisebox{-6pt}{$v$}\quad\raisebox{1pt}{$r$}} & 
     0 & 1 & 2 & \dots &  M-1 \\ \hline
 0   & 01     & 10     & 2(N-1)     & \dots & (M-1)(M+2)  \\
 1   & 12     & 21     & 30         & \dots & M(M+3)  \\
 2   & 23     & 32     & 41         & \dots & (M+1)(M+4) \\
 \dots & \dots & \dots & \dots      & \dots \\
 M-1 & (M-1)M & M(M-1) & (M+1)(M-2) & \dots & (N-2)1 
\end{array}
$$
  It can be easily observed that the index $p$ has $N-1$ values 
0,1,\dots,$N-2$ and the index $q$ has the same $p+1$ values for $p$ and 
$p'=N-2-p$. These values are $N-p$, $N-p+2$, \dots, $p$ in the first case
and $N-p+1$, $N-p+3$, \dots, $p+1$ for the representation $D^{l;10}$. 

To complete the discussion one has to consider representations $D^{l;k_11}$.
The irreducible bases of products $D^{l;k_11}\otimes D^{-l}$ have a form 
similar
to those given by Eqs.~(\ref{rep1}) and (\ref{rep2})
 \begin{eqnarray}
  \ket{w}_t^v&=&(-1)^\alpha\ket{s+v}_2^{01}\ket{w+2v}_-\,,  
      \label{rep3}\\
  \ket{w}_t^v&=&(-1)^\alpha\ket{s+v}_2^{11}\ket{w+2v+1}_-\,, 
   \label{rep4}
 \end{eqnarray}
 where $\alpha$ is an integer part of $(w+v)/M$. Since
$w+v=0,1,2,\dots,N+M-2$ then $(-1)^\alpha=-1$ for $w+v=M,M+1,\dots,N-1$. 
The same example as in the previous case ($N=6$, $l=1$) leads to the 
following bases
  \begin{eqnarray*}
  \left\{\ket{0}_2^{01}\ket{0}_-\;, \ket{1}_2^{01}\ket{1}\;,
    \ket{2}_2^{01}\ket{2}_-\;, -\ket{0}_2^{01}\ket{3}\;,
      -\ket{1}_2^{01}\ket{4}_-\;, -\ket{2}_2^{01}\ket{5}\right\}\;;\\
  \left\{\ket{1}_2^{01}\ket{2}_-\;, \ket{2}_2^{01}\ket{3}\;,
    -\ket{0}_2^{01}\ket{4}_-\;, -\ket{1}_2^{01}\ket{5}\;,
      -\ket{2}_2^{01}\ket{0}_-\;, \ket{0}_2^{01}\ket{1}\right\}\;;\\
  \left\{\ket{2}_2^{01}\ket{4}_-\;, -\ket{0}_2^{01}\ket{5}\;,
    -\ket{1}_2^{01}\ket{0}_-\;, -\ket{2}_2^{01}\ket{1}\;,
      \ket{0}_2^{01}\ket{2}_-\;, \ket{1}_2^{01}\ket{3}\right\}\;;\\
  \left\{\ket{0}_2^{11}\ket{1}_-\;, \ket{1}_2^{11}\ket{2}\;,
    \ket{2}_2^{11}\ket{3}_-\;, -\ket{0}_2^{11}\ket{4}\;,
      -\ket{1}_2^{11}\ket{5}_-\;, -\ket{2}_2^{11}\ket{0}\right\}\;;\\
  \left\{\ket{1}_2^{11}\ket{3}_-\;, \ket{2}_2^{11}\ket{4}\;,
    -\ket{0}_2^{11}\ket{5}_-\;, -\ket{1}_2^{11}\ket{0}\;,
      -\ket{2}_2^{11}\ket{1}_-\;, \ket{0}_2^{11}\ket{2}\right\}\;;\\
  \left\{\ket{2}_2^{11}\ket{5}_-\;, -\ket{0}_2^{11}\ket{0}\;,
    -\ket{1}_2^{11}\ket{1}_-\;, -\ket{2}_2^{11}\ket{2}\;,
      \ket{0}_2^{11}\ket{3}_-\;, \ket{1}_2^{11}\ket{4}\right\}\;.
  \end{eqnarray*}
  The formulae (\ref{rep1}) and (\ref{rep2}) have to be modified accordingly.
However, the elements $\ket{v+r}\ket{v-r}$ and $\ket{v+r}\ket{v-r+1}$, 
corresponding to $w=s=0$, always have coefficient +1, since $v<M$ and $w=0$
when one calcualtes $\alpha$ in Eqs.~(\ref{rep3}) and (\ref{rep4}). Therefore,
the set of indices $pq$ determined in the previous cases need not be changed.
If one wants replace a pair $pq$ by its counter-part $(p+M)(q+M)$, for example
$35$ by $02$ in the considered example, then a sign in the corresponding  
formula, (\ref{rep3}) or (\ref{rep4}), has to be changed, {\it i.e.}\
$\alpha$ is replaced by $\alpha+1$.

\section{Symmetrization of states} \label{sym}

In the previous section the relation between the basis $\ket{w}^{pq}_t$,
determined in \cite{trio} by a direct evaluation of a product of three 
representations
$D^l\otimes D^l\otimes D^{-l}$, and the basis $\ket{w}_t^{r,v;k_1k_2}$,
presenetd in this paper, is established.
Since the symmetrization of the basis $\ket{w}_t^{pq}$ and its relation
with a basis related to the product $D^l\otimes(D^l\otimes D^{-l})$ are
considered in the earlier paper \cite{trio}, then the problem sated in
the title of this section may be left out. However, having at hand
the basis $\ket{s}_2$ for pairs of identical particle it is natural and 
obvious to investiagte the symmetry properties related with the 
transposition of particles.

As in all above problems the case of odd $N=2M+1$ is quite easy. It follows
from Eq.~(\ref{oddbas}) that $r=0$ leads to symmetric states
  $$
   \ket0\ket0, \ket1\ket1,\dots, \ket{N-1}\ket{N-1}\,.
  $$
 The other $N^2-N$ vectors are grouped in $N-1=2M$ representations labelled
by $r=1,\dots,N-1$. To construct symmetric (antisymmetric) states it is 
enough to take combinations
  \begin{equation}
     \ket{s}_2^{r\pm} = \ket{s+r}\ket{s-r}\pm \ket{s-r}\ket{s+r}\,,
  \end{equation}
  where $r=1,2,\dots,M$ now.

In the case of even $N=2M$ the product $D^l\otimes D^l$ decomposes into
$M$ copies of four $M$-dimensional projective representations. Since
the symmetrization of states has a slightly different form in each
of these cases then they are considered separately.

The non-symmetrized basis of the representation $D^{l;00}$ is given
by Eq.~(\ref{ba00}) as
 $$
  \ket{s}_2^{r;00} = \ket{s+r}\ket{s-r}
      +\ket{M+s+r}\ket{M+s-r}\,.
  $$
 The case $r=0$ again gives $M$ symmetric states. To consider the other 
$M-1$ representations one has to check the parity of $M$. If $M=2\mu$,
then for $r=\mu$ we have $M-r=r$ and $M+r=N-r$, so in this case all
vectors are symmetric. The other vectors form symmetric and antisymmetric
states in the standard way and it also concerns the case of odd $M$. 

The `additional' symmetric state found above for $N=4\mu$ is `lost'
considering $D^{l;10}$. Its irreducible basis (\ref{ba10})
 $$
  \ket{s}_2^{r;10} = \ket{s+r}\ket{s-r+1}
      +\ket{M+s+r}\ket{M+s-r+1}
  $$
 have no symmetric states for $M=2\mu$, but for odd $M=2\mu-1$ and $r=\mu$
the above formula reads
 \begin{equation}
  \ket{s}_2^{\mu;10} = \ket{s+\mu}\ket{s-\mu+1}
      +\ket{s-\mu+1}\ket{s+\mu}\,,\label{symm}
 \end{equation}
 so these states are symmetric.

The similar considerations have to be performed for representations 
$D^{l;k_11}$ with special attention to the fact that 
 $$
  \ket{s-r}\ket{s+r} -\ket{M+s-r}\ket{M+s+r}
  =-(\ket{s+r'}\ket{s-r'}
      +\ket{M+s+r'}\ket{M+s-r'})\,,
  $$
where $r=0,1,\dots,M-1$ and $r'=M-r$.

\section{Final remarks}
  The presented considerations have shown that the parity of the BvK period
strongly influences the obtained results, whereas free trions 
behave in similar way as free electrons or holes and, therefore, their
states does not depend on the parity. Of course, coupling of three (in general,
$n$) identical partilces will lead to problems when $N=3M$ ($N=nM$, in a 
general case). It should be stressed that 
the degeneracy of obtained states is very high and there are
many possibilities to construct states $\ket{s}_2$ of pairs and $\ket{w}_t$ of
a three-particle systems. So we have to keep in mind that only a few 
possible choices of the irreducible basis has been presented in this paper.
For example, the formulae (\ref{oddbas})--(\ref{ba11}) and following them
Eqs.~(\ref{trioddbas}), (\ref{ba00r1}), (\ref{ba10r2}), are asymmetric
with respect to the transposition of identical particles. Symmetrization of
such states has been brielfy discussed in Sec.~\ref{sym}.  
In these simplified considerations there are
no interactions between trions or Landau level mixing and, moreover, the spin
or angular momentum numbers. Taking into account spins will allow to construct
states completely antisymmetric with respect to the permutational symmetry.
Such problem has been discussed lately by Dzyubenko {\it et al.} \cite{Dzy}
in the absence of a periodic potential $V(\mathbf{r})$, so
there is no discrete translational symmetry. The relations 
for the total angular moment projections 
obtained there are the same as those presented in this paper for indices
of vectors (taking into account the sign of charges). For example,
in Eq.~(\ref{trioddbas}) one obtains 
 $$
   w = (w+v+r)+(w+v-r)-(w+2v)\,.
 $$  
It is interesting
that Dzyubenko {\it et al.} obtained their results in the antisymmetric
gauge ${\mathbf A}=({\mathbf H}\times{\mathbf r})/2$, whereas in the presented
considerations the Landau gauge has been used. It confirms that the physical
properties are gauge-independent. On the other hand, the actual form of wave 
functions is not discussed here, but the relations between
representations and their product are taken into account only. These relations
are independent of the matrix representations and, similarly, the form of
resultant basis is independent of the function form: for a given BvK period
$N$ and any gauge irreducible projective representations are $N$-dimensional
and their action on basis vectors are similar (up to a factor system) 
\cite{brow,zak,jmp,fish}.

\end{document}